\documentstyle[preprint,aps,epsf]{revtex}
\tighten

\begin{document}

\preprint{\parbox[b]{3.3 cm} {NIKHEF 96-016\\hep-ph/9603428}}

\draft

\title{Higher-twist quark-mass contributions \\
to deep-inelastic scattering}
\author{D. Boer and R.D. Tangerman}

\address{National Institute for Nuclear Physics and High Energy Physics
(NIKHEF),\\ P.O. Box 41882, NL-1009~DB Amsterdam, The Netherlands }

\date{March 1996}

\maketitle

\begin{abstract}
In this letter we extend the factorization procedure of the deep-inelastic
hadron tensor, proposed by Qiu, to include non-zero quark masses.
The manifest gauge invariance of both soft and hard parts is preserved.
Using a so-called spurion to generate the quark-mass terms,
the simple parton-model interpretation is also kept.
The calculation of the deep-inelastic transverse-spin structure function $g_2$
is used to illustrate the algorithm.
\end{abstract}


The masses of the light quarks are small as compared to the typical
hadronic mass scales. Therefore, in most calculations of deep-inelastic
scattering (DIS) structure functions they are neglected.
In some cases, however, this is not allowed. A notorious example is
$g_2$ of a free quark target at tree-level~\cite{jaff91a}.
In that case the twist-three quark-mass term exactly cancels the twist-two
contribution, yielding $g_2=0$. In general, if the ratio $m/M$ is not
negligible, one is forced to keep the quark-mass contributions.

Let us recapitulate some historic facts.
In Ref.~\cite{poli80} Politzer suggests a diagrammatical
approach to unravel the power corrections in $1/Q$ for a wide class of
hard scattering processes.
This method was fully employed for the deep-inelastic scattering process
by Ellis, Furma\'nski, and Petronzio (EFP) in Ref.~\cite{EFP}, in which
they calculate the twist-four corrections to the unpolarized structure
functions. Thereto they factorize the diagrams in hard, i.e.\ perturbative, and
soft, i.e.\ non-perturbative, parts.
An essential ingredient in their calculation is the expansion
of the parton momenta around the target momentum direction, called the {\em
collinear expansion}. Using low-energy Ward identities they show that in all
matrix elements the gluon field combines with a derivative to precisely
the covariant derivative. Color gauge invariance is then manifest upon
inclusion of so-called link operators. Electromagnetic gauge invariance of the
hard scattering parts is not manifest in their approach. Matrix elements with a
definite parton number contribute at different orders in $1/Q$.

It was Qiu in Ref.~\cite{qiu90} who showed that, if one uses the concept of
the {\em special propagator\/}, one can factorize such that a complete
separation of the hard parts between different orders in $1/Q$ can be obtained.
Hence, the electromagnetic gauge invariance of these hard parts is manifest.
Moreover, only matrix elements with a fixed number of partons contribute to a
particular order in $1/Q$. This fact also enables a clear parton-model
interpretation of the higher-twist terms.

Both EFP and Qiu neglect quark masses. Our goal is to include them in such a
fashion that both color and electromagnetic gauge invariance remain manifest.
The basics of the method are as follows: in addition to the collinear
expansion, employed by EFP, we include a quark-mass expansion. Qiu's
factorization can be
extended as well. In order to preserve the parton interpretation we employ a
{\em spurion\/} particle~\cite{llew88}, which couples only to the fermions. The
spurions will generate the mass contributions. One maintains the three
advantages of combining EFP and Qiu's techniques, namely,
(1) the use of color gauge invariant matrix elements only,
(2) manifest electromagnetic gauge invariance of the hard scattering parts, and
(3) a clear parton interpretation at all times, also beyond leading order,
i.e., each order of power suppression means taking into account correlation
functions
with one parton more, where only the {\em good\/} fields~\cite{kogu70}
contribute.

The algorithm is best explained by applying it to a particular
case, therefore we look at polarized DIS up to and including order $1/Q$
(see Ref.~\cite{AEL} and the references therein).
The EFP approach in this context was first applied by Efremov and
Teryaev~\cite{efre84}.

The starting point is the familiar diagrammatic expansion of the forward
scattering amplitude $T^{\mu\nu}$ (we consider only quarks of one flavor for
the moment)
\begin{equation}
T^{\mu\nu}= \int d^{4}k \, \text{Tr} \left[ S^{\mu\nu} (k;m) \Gamma (k) \right]
+ \int d^{4}k_1  d^{4}k_2 \, \text{Tr} \left[ S^{\mu\nu}_{\alpha} (k_1,k_2;m)
\Gamma^{\alpha}_{A} (k_1,k_2) \right] + \ldots,
\label{expansion}
\end{equation}
keeping only the terms contributing up to order $1/Q$.
Here, $S$ and $\Gamma$ are the hard and soft scattering parts, respectively.
We will not consider QCD corrections, so the hard parts consist of the forward
parton-photon scattering tree graphs which are one-particle-irreducible after
shrinking the two photon vertices to a point. We indicate explicitly their
dependence
on the parton momenta and masses, but not on the photon momentum $q$
($q^2=-Q^2$). The soft parts are defined as
\begin{eqnarray}
\Gamma_{ij}(k)& = &\int \frac{d^{4}z}{(2\pi)^4} e^{ik\cdot z}
\langle \, P,S|T\left[\overline{\psi}_j (0) \psi_i (z)\right] | P,S\, \rangle ,
\\
\Gamma^{\alpha}_{Aij}(k_1,k_2)&=&\int \frac{d^{4}z}{(2\pi)^4}
\frac{d^{4}z'}{(2\pi)^4} e^{ik_1\cdot z} e^{i(k_2-k_1)\cdot z'} \langle \,
P,S|T\left[\overline{\psi}_j (0) g A^{\alpha}(z') \psi_i (z)\right]| P,S \,
\rangle .
\label{Gammaa}
\end{eqnarray}
Note that we have included a color identity and $g$ times $t^a$, respectively,
from the hard into the soft parts. Thereby the hard parts effectively
become QED graphs with unit charge~\cite{EFP}.

The nucleon is characterized by its momentum, satisfying $P^2=M^2$, and its
spin vector, satisfying $S\cdot P=0$ and $S^2=-1$.
We make a Sudakov decomposition of the relevant vectors with respect to two
light-like vectors $p$ and $n$, satisfying $p\cdot n=1$,
\begin{eqnarray}
q&=&-x_{\text{B}} p+\frac{Q^2}{2x_{\text{B}}}n,\\
P&=&p+\frac{M^2}{2}n,\\
S&=&\frac{\lambda}{M}\left(p-\frac{M^2}{2}n\right)+S_T,\\
k&=&x p+\frac{k^2-k_T^2}{2x}n+k_T.
\end{eqnarray}
Here, $x_{\text{B}}=Q^2/(2P\cdot q)$ is the Bjorken variable, $\lambda$ is the
nucleon's
helicity, $S_T$ its transverse spin. Also, $x$ is the quark's longitudinal
momentum, $k_T$ its transverse momentum. Since target mass corrections are
$1/Q^2$ suppressed, we ignore them.

The next step is performing a collinear and mass expansion of the hard parts
(keeping only relevant terms):
\begin{eqnarray}
S^{\mu\nu} (k;m) &= &S^{\mu\nu} (xp;0) + (k-xp)^{\alpha} \left.\frac{\partial
S^{\mu\nu} (k;m)}{\partial k^{\alpha}}
\right|_{\stackrel{\scriptstyle k=xp}{\scriptstyle m=0}}
+ m \left. \frac{\partial S^{\mu\nu} (k;m)}{\partial m}
\right|_{\stackrel{\scriptstyle k=xp}{\scriptstyle m=0}}
+ \ldots,\\
S^{\mu\nu}_{\alpha} (k_1,k_2;m) &=& S^{\mu\nu}_{\alpha} (x_1 p, x_2 p;0) +
\ldots.
\label{col}
\end{eqnarray}
We use the following Ward identities to rewrite the partial derivatives:
\begin{eqnarray}
\frac{\partial S^{\mu\nu} (k;m)}{\partial k^{\alpha}}&=& S^{\mu\nu}_{\alpha}
(k,k;m),\label{WI}\\
\frac{\partial S^{\mu\nu} (k;m)}{\partial m} &=& S^{\mu\nu}_{\text{spur}}
(k,k;m),\label{spur}
\end{eqnarray}
where the right-hand side of Eq.~(\ref{WI}) is effectively obtained from
$S^{\mu\nu} (k;m)$
by insertion of a zero-momentum gluon, with coupling $i\gamma^{\rho}_{ij}$
to the fermions.
The right-hand side of Eq.~(\ref{spur}) follows from the insertion of a
zero-momentum (scalar) spurion~\cite{llew88} (denoted by a dashed line) which
couples to
the fermions through the vertex $-i{\bf 1}_{ij}$.
Inserting the Ward identities, one arrives at
\begin{eqnarray}
T^{\mu\nu} &=& \int dx \, \text{Tr} \left[ S^{\mu\nu} (xp;0) \Gamma (x) \right]
+\int d x_1 d x_2 \, \text{Tr} \left[S^{\mu\nu}_{\alpha} (x_1p,x_2p;0)
{\omega^\alpha}_{\beta} \Gamma^{\beta}_{D} (x_1,x_2)\right]\nonumber\\
&&+ \int d x_1 d x_2 \, \text{Tr} \left[S^{\mu\nu}_{\text{spur}} (x_1p,x_2p;0)
\Gamma_{m} (x_1,x_2) \right]
+ \ldots, \label{efp}
\end{eqnarray}
where the projector
${\omega^\alpha}_{\beta}={g^\alpha}_{\beta}-p^{\alpha} n_{\beta}$.
Note that $A^{\alpha}={\omega^\alpha }_{\beta} A^{\beta}$ if one
uses the light-cone gauge $n\cdot A =0$. The soft parts read
($iD= i\partial+ g A$)
\begin{eqnarray}
\Gamma_{ij} (x) & = & \int \frac{d \lambda}{2\pi}e^{i\lambda x}\langle \,
P,S|T\left[\overline{\psi}_j (0) \psi_i(\lambda n)\right]| P,S \, \rangle ,\\
\Gamma_{Dij}^{\alpha} (x_1,x_2) & = & \int \frac{d \lambda}{2\pi} \frac{d
\eta}{2\pi} e^{i\lambda x_1} e^{i\eta (x_2 -x_1)} \langle \,
P,S|T\left[\overline{\psi}_j (0) iD^{\alpha}(\eta n) \psi_i(\lambda n)\right]
|P,S \, \rangle ,\\
\Gamma_{mij} (x_1,x_2) & = &  \int \frac{d \lambda}{2\pi} \frac{d \eta}{2\pi}
e^{i\lambda x_1} e^{i\eta (x_2 -x_1)} \langle \, P,S|T\left[\overline{\psi}_j
(0) m
\psi_i(\lambda n)\right]| P,S \, \rangle \nonumber\\
&=&m \,\delta(x_2-x_1)\,\Gamma_{ij} (x_1).
\label{blobs}
\end{eqnarray}
One can restore color gauge invariance of these definitions by introducing
path-ordered exponentials called link operators. In the gauge $n\! \cdot \! A
=0$, these links become unity, if the paths are chosen along the $n$-direction.
In Ref.\ \cite{qiu91b} it is shown that the gluons in the p-direction, i.e., of
the form $(n\cdot A)\, p$, can be pulled into the two-parton amplitude,
generating a link operator, showing that the gauge $n\! \cdot \! A =0$ is
inessential and color gauge invariance can always be achieved.

The first term in Eq.~(\ref{efp}) contributes to order $(1/Q)^0$ {\em and
higher},
whereas the latter terms contribute to order $(1/Q)^1$ {\em and higher}.
The new ingredient due to Qiu~\cite{qiu90} is basically the splitting of
the different terms into parts which contribute at a specific
order in $1/Q$.
Let us illustrate how this can be done. The Dirac trace in the first term in
Eq.~(\ref{efp}) can be Fierz decomposed according to
\begin{equation}
\text{Tr} \left[ S^{\mu\nu} (xp;0) \Gamma (x) \right] =  \case{1}{4} \text{Tr}
\left[ S^{\mu\nu} (xp;0) \gamma_{\rho}  \right] \text{Tr} \left[\gamma^{\rho}
\Gamma (x) \right]
+\case{1}{4} \text{Tr} \left[ S^{\mu\nu} (xp;0) \gamma_{5} \gamma_{\rho}
\right] \text{Tr} \left[\gamma^{\rho}\gamma_{5} \Gamma (x) \right],
\label{Fierz}
\end{equation}
where we used the fact that in the hard part $m=0$,
such that chirality is conserved.
The vector term contributes to the
unpolarized scattering only, so we discard it.
We make a Sudakov decomposition of the axial-vector projection of the soft
part,
\begin{equation}
\text{Tr} \left[\gamma^{\rho}\gamma_{5} \Gamma (x) \right] = p^{\rho} \text{Tr}
\left[\mbox{$\not\! n\,$}\gamma_{5} \Gamma (x) \right] + \text{Tr}
\left[\gamma^{\rho}_T \gamma_{5} \Gamma (x) \right] + n^{\rho} \text{Tr}
\left[\mbox{$\not\! p\,$}\gamma_{5} \Gamma (x) \right],
\label{Sud}
\end{equation}
where
$\gamma_{T}^\rho=g_T^{\rho\sigma}\gamma_{\sigma}
\equiv (g^{\rho\sigma}-p^\rho n^\sigma -n^\rho p^\sigma)\gamma_{\sigma}$.
The only dimensionful quantity in the hard part is $Q$, whereas the soft
parts are assumed not to contain large scales.
This, in combination with the fact that $p$ has dimension $1$, and $n$
dimension $-1$, leads to the observation that the first term is the leading
one, while the second and third are $1/Q$ and $1/Q^2$ suppressed, respectively,
so we discard the last one.

\begin{figure}
\vspace{2.5cm}
\hspace{3.1cm}
\epsfbox{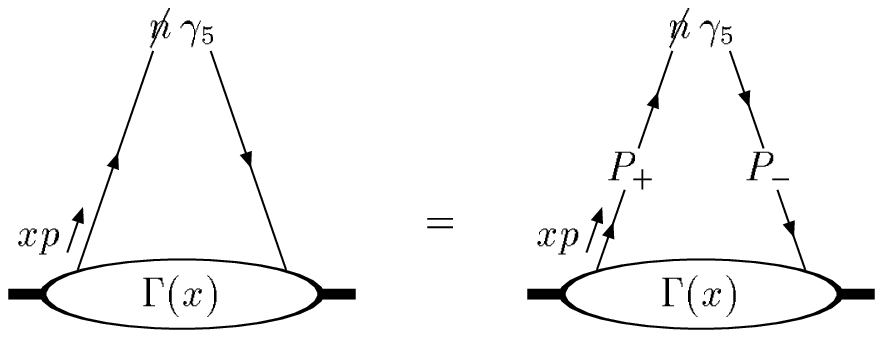}
\vspace{1cm}
\caption{\label{fig:trace1}The leading axial-vector projection of $\Gamma(x)$.}
\end{figure}

\begin{figure}
\vspace{2cm}
\hspace{0.7cm}
\epsfbox{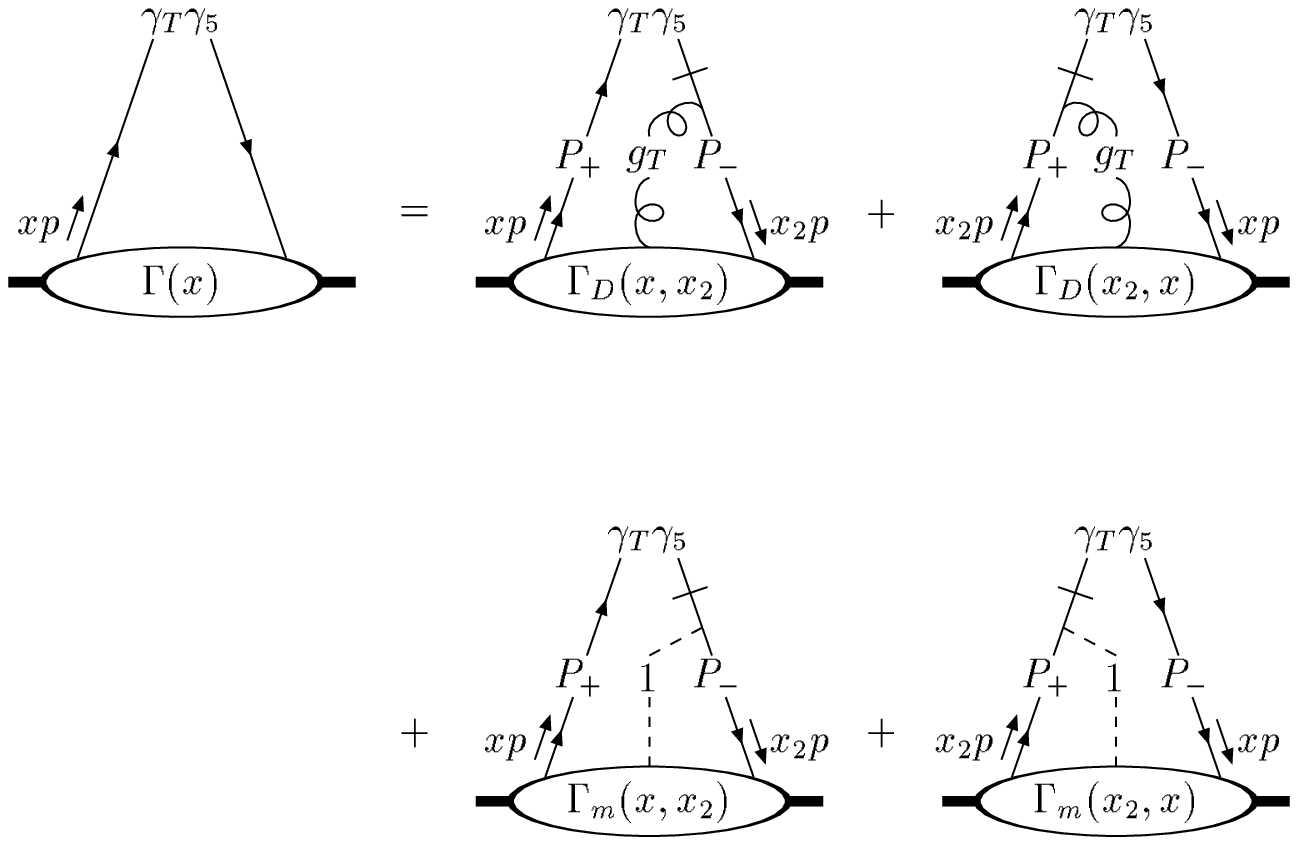}
\vspace{1cm}
\caption{\label{fig:trace2}The sub-leading axial-vector projection of
$\Gamma (x)$.}
\end{figure}

Consider the leading $\mbox{$\not\! n\,$}\gamma_{5}$ trace first.
Introducing the projectors $P_{+} = \mbox{$\not\! p\,$} \mbox{$\not\! n\,$}/2$
and
$P_{-} = \mbox{$\not\! n\,$}\mbox{$\not\! p\,$}/2$, which project on `good' and
`bad' quark fields,
respectively~\cite{kogu70}, and using the relations
\begin{equation}
\mbox{$\not\! n\,$}=P_-\mbox{$\not\! n\,$}=\mbox{$\not\! n\,$}P_+,\label{aap}
\end{equation}
it can be diagrammatically represented as in Fig.~\ref{fig:trace1}.
Consider next the trace of $\Gamma(x)$ with $\gamma_T\gamma_{5}$, which we
write as
$(\mbox{$\not\! p\,$}\gamma_{T}\gamma_{5}\mbox{$\not\! n\,$}+\mbox{$\not\!
n\,$}\gamma_{T}\gamma_{5}\mbox{$\not\! p\,$})/2$.
Again, we can multiply the \mbox{$\not\! n\,$}
with a $P_+$ on the right, or with a $P_-$ on the left.
For the \mbox{$\not\! p\,$} we use the following relations,
which follow from the equations of motion $i\mbox{$\,\not\!\!
D$}\psi=m\psi$~\cite{poli80},
\begin{eqnarray}
\Gamma (x) \mbox{$\not\! p\,$}& = & \int dx_2 \,  \left[ \Gamma^{\beta}_{D}
(x,x_2)
{\omega^\alpha}_{\beta} (i\gamma_{\alpha}) +\Gamma_{m} (x, x_2) (-i{\bf
1})\right] \frac{i\mbox{$\not\! n\,$}}{2x} \mbox{$\not\! p\,$},\label{spec}\\
\mbox{$\not\! p\,$}\Gamma (x) & = & \mbox{$\not\! p\,$}\frac{i\mbox{$\not\!
n\,$}}{2x}\int dx_2 \,
\left[ (i\gamma_{\alpha}){\omega^\alpha}_{\beta}\Gamma^{\beta}_{D} (x_2,x)
+(-i{\bf 1})\Gamma_{m} (x_2, x) \right]
\end{eqnarray}
Finally, we use
\begin{eqnarray}
{\omega^\alpha}_{\beta} \gamma_{\alpha}\mbox{$\not\! n\,$}&=&
P_-\,{{g_T}^\alpha}_{\beta} \gamma_{\alpha}\mbox{$\not\! n\,$},\\
\mbox{$\not\! n\,$}\gamma_{\alpha}{\omega^\alpha}_{\beta}&=&
\mbox{$\not\! n\,$}\gamma_{\alpha}{{g_T}^\alpha}_{\beta} P_+,
\end{eqnarray}
along with Eq.~(\ref{aap}), to project out the good quark and good (transverse) 
gluon fields.
Putting the pieces together, we can write the trace diagrammatically as
a sum of four diagrams; two gluon and two spurion diagrams.
They are depicted in Fig.~\ref{fig:trace2}.
We have used the {\em special propagator}~\cite{kogu70,qiu90}
\begin{equation}
\begin{picture}(50,20)(0,5)
\put(5,5){\line(1,0){30}}
\put(20,0){\line(0,1){10}}
\put(5,12){\vector(1,0){10}}
\put(3,5){\makebox(0,0)[r]{$\scriptstyle j$}}
\put(37,5){\makebox(0,0)[l]{$\scriptstyle i$}}
\put(5,15){\makebox(0,0)[bl]{$\scriptstyle k$}}
\end{picture}
=\frac{i\mbox{$\not\! n\,$}_{ij}}{2k\cdot n}.
\end{equation}
The essence of the above derivation is that an \mbox{$\not\! n\,$} pulls out a
good
quark (or anti-quark) from the soft part, whereas a \mbox{$\not\! p\,$} pulls
out a special propagator, a good quark, along with a good gluon or
(for non-zero quark mass) spurion. Note that for the last trace in
Eq.~(\ref{Sud}) this implies the inclusion of four-parton amplitudes, leading
to $1/Q^2$ suppression.

In order to undo the Fierz decomposition, Eq.~(\ref{Fierz}), we use the
following relations:
\begin{eqnarray}
P_{-} p^\rho \mbox{$\not\! n\,$} P_{+} &=& P_{-} \gamma^{\rho} P_{+}, \\
P_{-} \mbox{$\not\! n\,$} \gamma^{\rho}_T  P_{+}&=& P_{-} \mbox{$\not\! n\,$}
\gamma^{\rho} P_{+} ,\\
P_{-} \gamma^{\rho}_T \mbox{$\not\! n\,$} P_{+} &=& P_{-} \gamma^{\rho}
\mbox{$\not\! n\,$} P_{+}.
\end{eqnarray}

\begin{figure}
\hspace{0.9cm}
\epsfbox{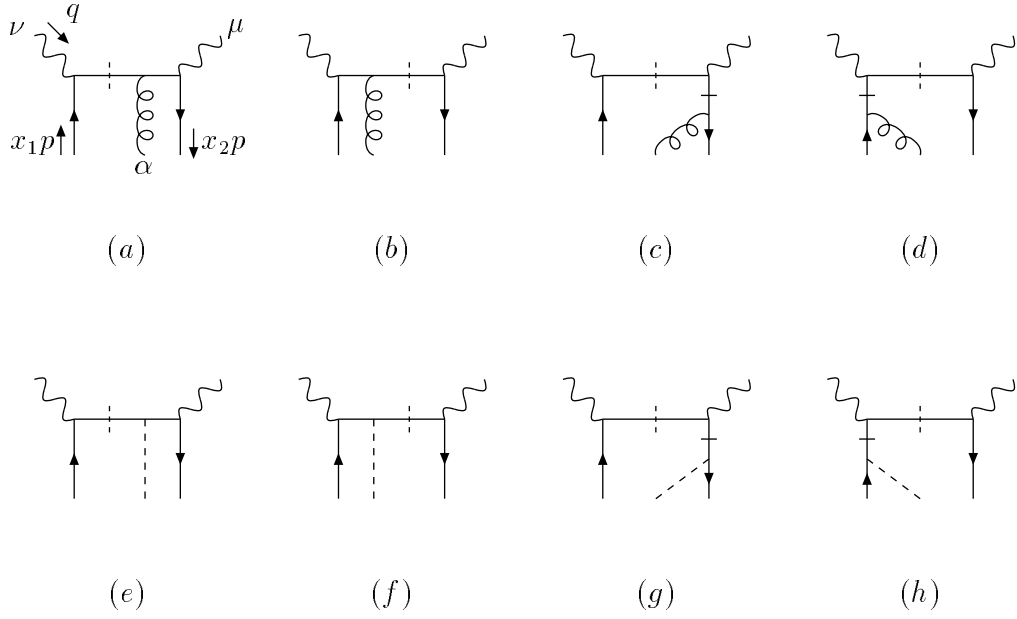}
\vspace{1cm}
\caption{\label{fig:hard}Uncrossed twist-three gluonic [$(a)$-$(d)$]
and spurionic [$(e)$-$(h)$] diagrams.}
\end{figure}

Returning to the second and third terms in Eq.~(\ref{efp}), from
dimensional arguments one can again infer that only good fields contribute
at leading order. So we may include projectors in between the hard and
soft parts.
The crucial step in the Qiu factorization is to absorb everything above the
projectors into the upper, hard, part (see Figs.~\ref{fig:trace1}
and~\ref{fig:trace2}).
For the leading order this has no consequences.
At sub-leading order, however, the modified hard gluonic and spurionic
parts read
\begin{eqnarray}
H_{\alpha}^{\mu\nu}(x_1p,x_2p;0) &=& S^{\mu\nu}_{\alpha} (x_1p,x_2p;0) +
i\gamma_{\alpha}\frac{i\mbox{$\not\! n\,$}}{2x_1} S^{\mu\nu} (x_1p;0) +
S^{\mu\nu} (x_2p;0) \frac{i\mbox{$\not\! n\,$}}{2x_2}i\gamma_{\alpha},
\label{Ha}\\
H_{\text{spur}}^{\mu\nu}(x_1p,x_2p;0) &=& S^{\mu\nu}_{\text{spur}}
(x_1p,x_2p;0) +(-i{\bf 1})
\frac{i\mbox{$\not\! n\,$}}{2x_1} S^{\mu\nu} (x_1p;0)
+ S^{\mu\nu} (x_2p;0) \frac{i\mbox{$\not\! n\,$}}{2x_2}(-i{\bf 1}). \label{Hs}
\end{eqnarray}
The treatment of the spurionic part is the new result of this paper. The
leading and sub-leading forward scattering amplitudes become
\begin{eqnarray}
T^{\mu\nu}_{\text{twist-2}} &=&
\int dx \, \text{Tr} \left[ P_{-} S^{\mu\nu} (xp;0) P_{+} \Gamma (x)
\right],\label{LO}\\
T^{\mu\nu}_{\text{twist-3}} &=&
\int dx_1 \, dx_2 \, \text{Tr} \left[ P_{-} H_{\alpha}^{\mu\nu}(x_1p,x_2p;0)
P_{+}
{{g_T}^\alpha}_{\beta} \Gamma^{\beta}_{D} (x_1, x_2) \right]\nonumber\\
&& +\int dx_1 \, dx_2 \, \text{Tr} \left[ P_{-}
H_{\text{spur}}^{\mu\nu}(x_1p,x_2p;0)
P_{+} \Gamma_{m} (x_1, x_2) \right],\label{NLO}
\end{eqnarray}
which are of a specific order in $1/Q$. That is, `twist-$t$' contributes only
at order $Q^{2-t}$. Also, $t$ always equals the number of partons connecting
the hard and soft parts. To arrive at the DIS hadronic tensor,
$W^{\mu\nu}=\text{Disc}[T^{\mu\nu}]/(4M\pi i)$, one has to cut the hard
diagrams.
The uncrossed twist-three cut diagrams are depicted in Fig.~\ref{fig:hard}.
Explicitly, one finds for the (projected) discontinuities (displaying only the
antisymmetric parts in $\mu\leftrightarrow\nu$):
\begin{eqnarray}
\text{Disc}\left[P_-S^{\mu\nu}(x p;0)P_+\right]&=&
\pi i e^2\delta(x-x_{\text{B}})\;i{\epsilon_T}^{\mu\nu}\mbox{$\not\!
n\,$}\gamma_{5},\label{haver}\\
\text{Disc}\left[P_-H^{\mu\nu}_{\alpha}(x_1 p,x_2 p;0)P_+
{{g_T}^{\alpha}}_{\beta}\right]
&=&\frac{\pi i e^2}{Q^2}i\epsilon^{\mu\nu\rho\sigma}
q_\rho\Bigl\{i{\epsilon_T}_{\sigma\beta}\mbox{$\not\! n\,$}
\left[\delta(x_1-x_{\text{B}})-\delta(x_2-x_{\text{B}})\right]\nonumber\\
&&\qquad\qquad +{g_T}_{\sigma\beta}\mbox{$\not\! n\,$}\gamma_{5}
\left[\delta(x_1-x_{\text{B}})+\delta(x_2-x_{\text{B}})\right]
\Bigr\},\label{harda}\\
\text{Disc}\left[P_-H^{\mu\nu}_{\text{spur}}(x_1 p,x_2 p;0)P_+\right]
&=&\frac{\pi i e^2}{Q^2}i\epsilon^{\mu\nu\rho\sigma}q_\rho\mbox{$\not\! n\,$}
\gamma_{T\sigma}\gamma_{5}\left[\delta(x_1-x_{\text{B}})
+\delta(x_2-x_{\text{B}})\right].\label{gort}
\end{eqnarray}
where $\epsilon_T^{\mu\nu}=\epsilon^{\mu\nu\rho\sigma}p_\rho n_\sigma$.
The calculation of these discontinuities is especially simple
since the partons are massless, the effects of non-zero $m$ being already taken
into account.
At this point the electromagnetic gauge invariance is manifest. That is,
contraction of the modified hard parts with $q$ gives zero identically.
Compare this with the much more elaborate demonstration of gauge invariance in
Ref.~\cite[App. C]{AEL}.

For the soft parts we use the following parametrizations~\cite{jaff91b}
\begin{eqnarray}
\Gamma(x)&=&\frac{1}{2}g_1(x)\,\lambda\gamma_{5}\mbox{$\not\! p\,$}
+\frac{1}{2}h_1(x)\,\gamma_{5}\mbox{$\not\! S$}_T\mbox{$\not\! p\,$}
+\frac{M}{2}g_T(x)\,\gamma_{5}\mbox{$\not\! S$}_T+\ldots,\label{wal}\\
\Gamma^{\alpha}_{D}(x_1,x_2)&=&\frac{M}{2}G(x_1,x_2)\,i\epsilon_T^{\alpha
\beta} S_{T \beta} \mbox{$\not\! p\,$} + \frac{M}{2}\tilde{G}(x_1,x_2)\,
S_{T}^\alpha \gamma_{5} \mbox{$\not\! p\,$} + \ldots,\\
\Gamma_{m}(x_1,x_2)&=&\frac{M}{2}H_1(x_1,x_2)\,\gamma_{5}\mbox{$\not\!
S$}_T\mbox{$\not\! p\,$} + \ldots,
\label{sloot}\end{eqnarray}
where we listed only the distribution functions relevant for the process under
consideration.
The functions $\tilde{G}$ and $H_1$ are symmetric, $G$ is antisymmetric
under exchange of the two arguments.
The functions are not all independent. One has the following relations
\begin{eqnarray}
2x_1 g_T(x_1)&=&\int dx_2 \Bigl[G(x_1,x_2)-G(x_2,x_1)+\tilde{G}(x_1,x_2)+
\tilde{G}(x_2,x_1)\nonumber\\
&&\qquad\quad +H_1(x_1,x_2)+H_1(x_2,x_1)\Bigr],\label{eom}\\
H_1(x_1,x_2)&=&\frac{m}{M}\delta(x_2-x_1)h_1(x_1),
\end{eqnarray}
where the first equation follows from the equations of motion, the second from
Eq.~(\ref{blobs}). Contracting the hard parts, Eqs.~(\ref{haver})-(\ref{gort}),
with the soft parts, Eqs.~(\ref{wal})-(\ref{sloot}), gives for
the antisymmetric hadron tensor
\begin{eqnarray}
W^{A\mu\nu}_{\text{twist-2}}
&=&\frac{e^2}{2M}\,\lambda i\epsilon_T^{\mu\nu} g_1(x_{\text{B}}),\\
W^{A\mu\nu}_{\text{twist-3}}
&=&\frac{e^2}{2P\cdot q}i\epsilon^{\mu\nu\rho\sigma} q_\rho S_{T \sigma}
g_T(x_{\text{B}}),
\label{complete}
\end{eqnarray}
where we used Eq.~(\ref{eom}) to eliminate the two-argument functions.
Comparing with the standard expression in terms of structure functions
\begin{equation}
W^{A\mu\nu}=\frac{1}{P\cdot q}i\epsilon^{\mu\nu\rho\sigma}q_{\rho} \left[
S_\sigma g_1(x_{\text{B}},Q^2) + \left( S_\sigma - \frac{S\cdot  q}{P\cdot  q}
P_\sigma \right) g_2(x_{\text{B}},Q^2) \right],
\end{equation}
we identify the tree-level structure functions $g_1(x_{\text{B}},Q^2)= \sum_a
(e_{a}^2/2) \left[g_1^a(x_{\text{B}})+g_1^a(-x_{\text{B}})\right]$ and
$g_1(x_{\text{B}},Q^2)+g_2(x_{\text{B}},Q^2)= \sum_a (e_{a}^2/2)
\left[g_T^a(x_{\text{B}})+g_T^a(-x_{\text{B}})\right]$,
where we reinstalled the flavor sum and added the crossed (anti-quark)
diagrams.
These results are in agreement with the standard EFP-type of
calculations~\cite{efre84,ratc86,AEL}. It is the way they are derived that
makes
the difference. The algorithm can be summarized as follows. To obtain the
regular contributions to the order $Q^{2-t}$ DIS hadron tensor,
following Qiu~\cite{qiu90}, one has to (1) write down all possible forward
2-photon $t$-parton cut diagrams,
where partons can be quarks, antiquarks or gluons; (2) replace all propagators
that do not lie between the photon vertices by special propagators;
(3) project out the good fields\footnote{These
projections pick out one specific order in $1/Q$. They were not used
in Ref.~\cite{qiu90}.} and couple them with $t$-parton soft matrix elements.
The quark-mass terms follow in the same way, except that one (or more) of the
gluon legs that connect hard and soft parts is replaced by a spurion. In the
hard parts the partons are kept massless.

We have also applied the method to one-hadron-production in
e$^+$e$^-$-annihilation, which essentially only differs from the above by the
fact that the photon momentum is now time-like and the distribution functions
have
to be replaced by fragmentation functions (indicated with a hat, e.g.,
$\hat{g}_T^a$). We can again express the sub-leading structure function
(denoted by $\hat{g}_1+\hat{g}_2$) in terms of $\hat{g}^a_T$ only, that is,
without explicit mass terms.
This result differs from Eq.~(38) of Ref.~\cite{lu96}, in which Qiu's
factorization is also applied, by quark-mass terms (accompanied by the
fragmentation function $\hat{h}_1^a$). Our claim is that these terms should not
be present.

In this letter we have extended the formalism by Qiu, leading to a
factorization of hard and soft
scattering parts in a manifestly gauge-invariant way, to include non-zero quark
masses. We have illustrated the method for polarized DIS at sub-leading order.
Another interesting application would be the calculation of the quark-mass
contributions at twist-four in unpolarized DIS~\cite{lee94}. Also, one should
investigate whether the formalism can be extended to include QCD corrections.
Work in this direction is in progress.

\vspace{1cm}
We thank P.J. Mulders for useful comments.
This work was supported by the Foundation for Fundamental Research on Matter
(FOM) and the National Organization for Scientific Research (NWO).

\end{document}